\documentclass[usenatbib]{mnras}

\usepackage{newtxtext,newtxmath}

\usepackage{ae,aecompl}

\usepackage{graphicx}	
\usepackage{amsmath}	
\usepackage{amssymb}	
\usepackage{multicol}
\usepackage{hyperref}

\title[THOR and GLOSTAR]{Characterization of unresolved and unclassified sources detected in radio continuum surveys of the Galactic plane}

\author[A Chakraborty et al.]{
Arnab Chakraborty,$^{1}$\thanks{E-mail: arnab.phy.personal@gmail.com}
Nirupam Roy, $^{2}$
Y.Wang, $^{3}$
Abhirup Datta,$^{1}$
H. Beuther, $^{3}$
\newauthor
S.-N.X. Medina, $^{4}$
K. M. Menten,$^{4}$
J. S. Urquhart, $^{5}$
A. Brunthaler,$^{4}$
and S. A. Dzib $^{4}$
\\
$^{1}$Discipline of Astronomy, Astrophysics and Space Engineering, Indian Institute of Technology Indore, Indore 453552, India\\
$^{2}$Department of Physics , Indian Institute of Science,Bangalore 560012,India\\
$^{3}$Max Planck Institute for Astronomy, Königstuhl 17, 69117 Heidelberg, Germany \\
$^{4}$Max-Planck-Institut für Radioastronomie, Auf dem Hügel 69, 53121 Bonn, Germany\\
$^{5}$Centre for Astrophysics and Planetary Science, University of Kent, Canterbury, CT2 7NH, UK\\
}
\date{Accepted XXX. Received YYY; in original form ZZZ}

\pubyear{2015}

\begin{document}
\label{firstpage}
\pagerange{\pageref{firstpage}--\pageref{lastpage}}
\maketitle

\begin{abstract}
  The continuum emission from 1 to 2 GHz of  The HI/OH/Recombination line survey of the inner Milky Way (THOR) at $\lesssim18\arcsec$ resolution covers $\sim 132$ square degrees of the Galactic plane and detects 10387 sources. Similarly, the first data release of the Global View of Star Formation in the Milky Way (GLOSTAR)  surveys covers $\sim 16$ square degrees of the Galactic plane from 4-8 GHz at $18\arcsec$ resolution and detects 1575 sources. However, a large fraction of the unresolved discrete  sources detected in these radio continuum surveys of the Galactic plane remain unclassified.  Here, we study the Euclidean-normalized differential source counts of unclassified and unresolved sources detected in these surveys and compare them with simulated  extragalactic radio source populations as well as previously  established source counts. We find that the  differential source counts for THOR and GLOSTAR surveys  are in excellent agreement with both simulation and previous observations. We also estimate the angular two-point correlation function of unclassified and unresolved sources detected in  THOR survey. We find  a higher clustering amplitude in comparison with the Faint Images of the Radio Sky at Twenty-cm (FIRST) survey up to the angular separation of $5^{\circ}$. The decrease in angular correlation with increasing flux cut and the excellent agreement of clustering pattern of sources above 1 mJy with high $z$ samples ($z >0.5$) of the FIRST survey indicates that these sources might be high $z$ extragalactic compact objects. The similar pattern of one-point and two-point statistics of unclassified and compact sources with extragalactic surveys and simulations confirms the extragalactic origin of these sources.   
\end{abstract}




\section{Introduction}
Continuum surveys of  the Galactic plane at radio wavelength are an excellent way to study different source populations such as HII regions, planetary nebulae (PNe), radio stars etc \citep{Bihr2016,Beuther2016}. These surveys also help to understand different physical processes in the interstellar medium. There are several high-resolution surveys of the Galactic plane from near-infrared to mm wavelengths but only a few  at radio wavelength (see \citealt{Beuther2016} and references therein).  The HI/OH/Recombination line survey of the inner Milky Way (THOR) and the Global View of Star Formation in the Milky Way (GLOSTAR) are two such surveys of the Galactic plane at radio wavelengths in high-resolution with unprecedented sensitivity.

 THOR covers a large fraction of the first Galactic quadrant with the extended Karl G. Jansky Very Large Array (VLA) in C-configuration  at L-band from 1 to 2 GHz and detects 10387 sources \citep{Beuther2016,Wang2018}. \citet{Wang2018} have classified a subset of these sources as HII regions, pulsars,  X-ray sources, planetary nebulae, supernova remnants and extragalactic jets after comparing with multi-frequency catalogues.  GLOSTAR  covers the Galactic plane between -2$^{\circ} < l < 60^{\circ}$ and |$b$| < $1^{\circ}$, and then Cygnus X region from 76$^{\circ} < l < 83^{\circ}$ and -1<$b$<2 with the VLA in D- and B-configuration  at C-band from 4 to 8 GHz.  \citet{Medina2019} have analyzed  a portion of the D-array data and published the first source catalogue  consisting of 1575 sources. They have also identified the sources in the GLOSTAR survey by cross-correlating with different catalogues. However, a large number of sources remain unclassified in both surveys.  \citet{Wang2018} have found that the unclassified sources and identified Galactic  sources show different spatial and spectral index distributions. The Galactic sources are more concentrated in low longitude region and  near the Galactic mid-plane. Also, the spectral index distribution shows a peak around $\alpha \sim -1$ for unclassified sources in comparison with Galactic sources. However, they have also found a significant number of unclassified sources with $\alpha \geq 0$ \citep{Wang2018}.  \citet{Medina2019} have also found a similar distinctive behavior in spatial and spectral index distribution of unclassified sources in GLOSTAR catalogue over classified Galactic objects. The mean value of the spectral index of these unclassified sources $\sim -0.5 \pm 0.3$, which suggests that these sources are likely to be extragalactic origin \citep{Medina2019}. The clear difference in these distributions hinted that most of these unclassified sources are of extragalactic origin.  However, the exact origin of this large sample is still unknown and   there maybe are some exotic Galactic population. Here, we study  various properties of this sample to check the consistency with the extragalactic population. 

In this work, we study the one-point and two-point statistical properties of these unclassified and compact (or, unresolved) sources. We analyze the differential source counts as a function of flux density (one-point statistics) and compare the results with previous observations as well as with simulations. We have also estimated the angular two-point correlation function (ATCF: two-point statistics) of these sources using different flux density thresholds and compared with the FIRST survey.  This comparative  study of statistical properties of these sources allow us to identify whether they are of Galactic or extragalactic origin, which is also complementary to the previous findings of \citet{Wang2018} and \citet{Medina2019}.

\section{Data}
\label{Data}
In this section, we will briefly discuss about the radio catalogues used in this work to estimate the distribution of sources as a function of flux density  and angular clustering of sources.   

The details of  the observations, data reduction, accumulation of source catalogue of  THOR survey  are discussed and described in \citet{Bihr2016}, \citet{Beuther2016} and \citet{Wang2018}.  THOR survey covers the Galactic plane from $14.5^{\circ}$ to $67.25^{\circ}$ in Galactic longitude ($l$) and $\pm 1.25^{\circ}$ in Galactic latitude ($b$). The observation was done with VLA in C configuration at L band from 1 to 2 GHz divided into 8 continuum  spectral windows (SPWs).  \citet{Wang2018} have selected two least RFI affected SPWs (out of 6 SPWs)  centred at 1.82 GHz and 1.44 GHz, then smoothed them into the same resolution  ($ \lesssim 18\arcsec$; see Table.1 of \citealt{Wang2018}) and then made an average image. They have used BLOBCAT  \citep{Hales2012} to extract sources from this averaged image with $5\sigma$ as detection threshold and $2.6\sigma$ as flooding threshold. After removing the sidelobe artifacts they have assembled a catalogue  consisting of 10387 sources. However, a large number of sources, 9000 (86\%), of this catalogue remain unclassified. The nature of these sources are unknown. Out of these unclassified sources, 7800 (75\%) sources are unresolved. We have taken these unresolved and unclassified sources and estimated the one-point and two-point statistics. 
\begin{figure*}
\centering
\begin{tabular}{cc}
\includegraphics[width=\columnwidth,height=3in]{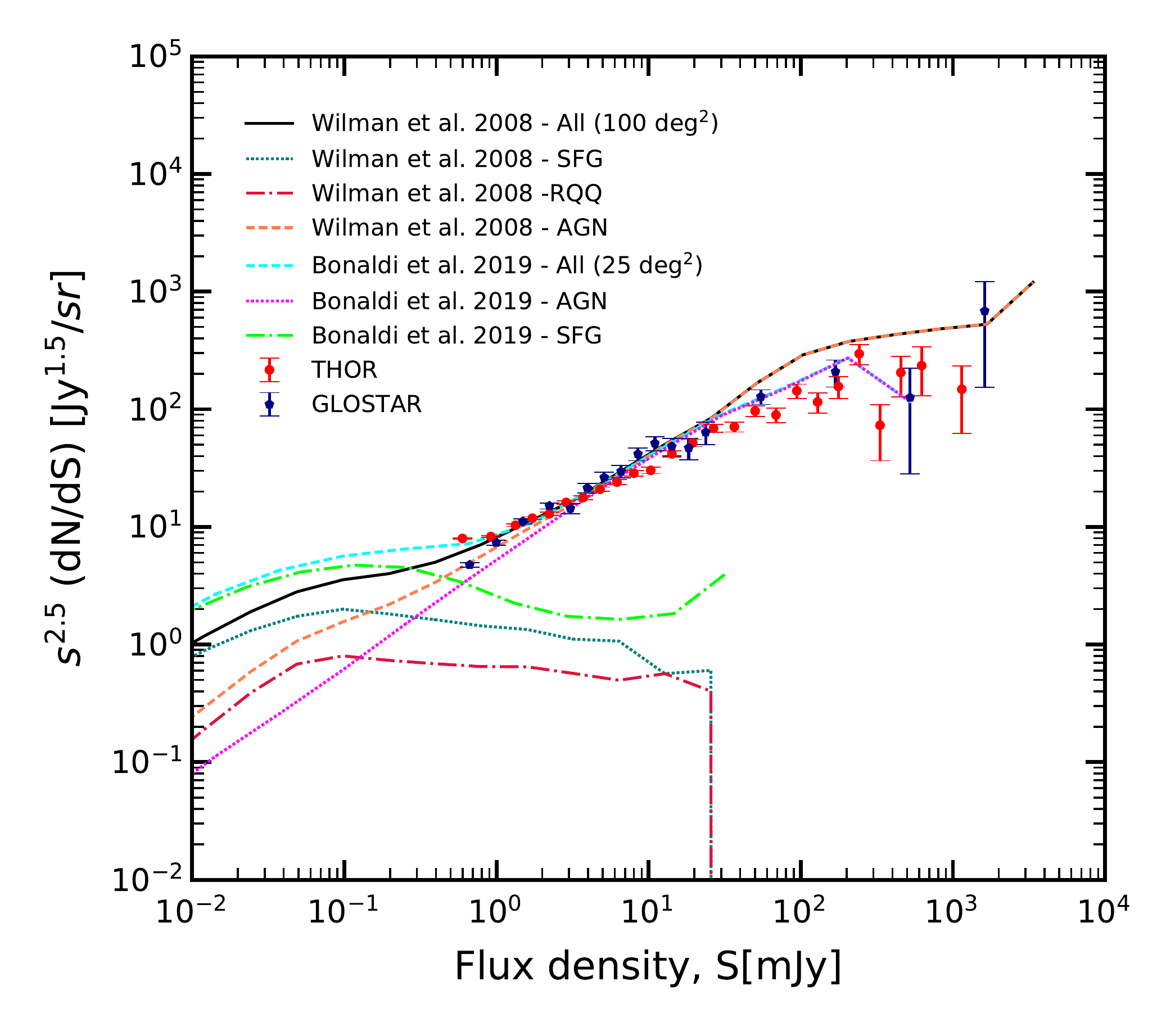}  & \includegraphics[width=\columnwidth,height=3in]{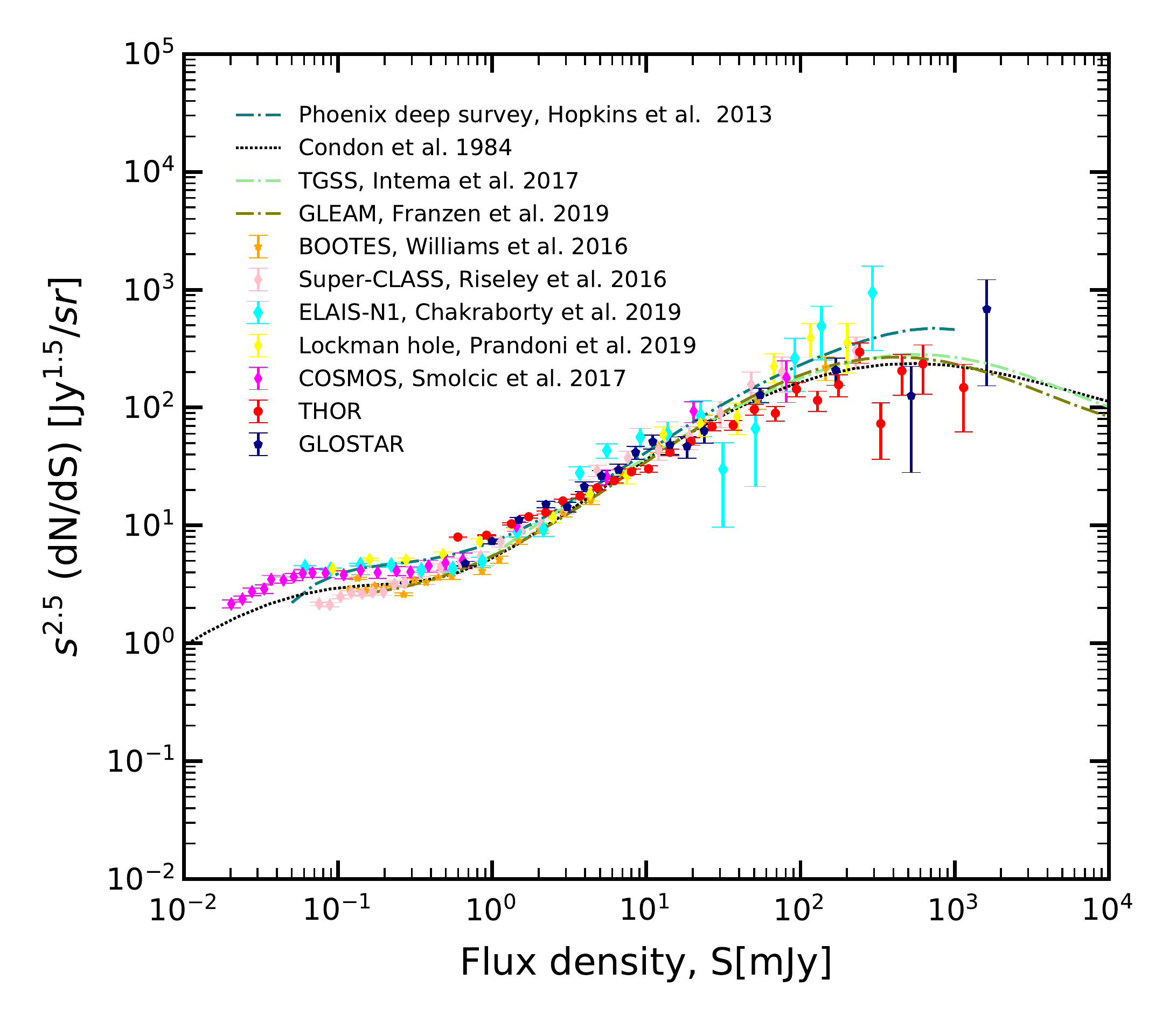} \\
  
\end{tabular}

\caption{The Euclidean-normalized differential source counts of unclassified and unresolved sources detected in  THOR and GLOSTAR survey compared with simulated radio sky (left panel) and previously observed source populations (right panel). For details of simulated catalogues and different observed source populations see text.}
\label{dn_ds}
\end{figure*}

         
    

The  GLOSTAR-VLA survey (Brunthaler et al., in prep.) was done using VLA  D- and B-configuration in C-band (4-8 GHz). \citet{Medina2019} have analyzed  a portion of the D-array data of this survey, which covers 16 $\mathrm{deg}^{2}$ region of the Galactic mid-plane spanning $28^{\circ} < l < 36^{\circ}$ and $|b| < 1^{\circ}$. They  made a combined mosaic image at an effective frequency of 5.8 GHz with $18\arcsec$ resolution and extracted the source catalogue from that image using BLOBCAT.  A total of 1575 sources  were included in the final catalogue after removing artifacts and identifying  multi-component sources. \citet{Medina2019} classified the sources as ionization fronts, pulsars, planetary nebulae, HII regions, stars, SNRs after cross-correlating with different catalogues. Here also a large number of  radio sources remain unclassified. \citet{Medina2019} have also classified the sources as the unresolved/compact if the ratio between integrated and peak flux density (Y-factor) of a object is  less than 1.2. There are 1284 (81\%) unresolved sources, out of which 1105 (70\%) sources are unclassified too. Similar to THOR,  we have only analyzed the statistical properties of  these unresolved and unclassified sources. 

\section{Differential source counts}
\label{source_counts}
There are extensive study of source counts as a function of flux density at high frequency as well as at low frequency.  It is well established that Euclidean-normalized differential source counts  at 1.4 GHz  show a flattening at $\sim$ 1 mJy corresponding to  a rise in  the source population of star-forming galaxies (SFGs) and radio-quiet AGNs (see \citealt{Zotti2010} for a detail review). We have estimated the Euclidean-normalized differential source counts of unclassified and unresolved sources detected in  THOR and GLOSTAR survey and compared our findings with previous observations. We have binned the integrated flux density of sources in logarithmic space and adjusted the bins in such a way that the highest flux density bin includes a minimum of 3 sources.  The raw source counts (N) in each bin are corrected for  image area detection fraction or the visibility area. We have estimated the fraction of area (f) over which a source with a given flux density can be detected (its visibility area) and weighted the raw source counts by the reciprocal of that fraction \citep{Windhorst1985}. The  corrected source counts ($\mathrm{N_{c}}$ = N/f) were then  divided by the total image area ($\Omega$ in steradians) and bin width ($\Delta S$ in Jy). This gives the differential source counts as a function of flux density. We have normalized the differential source count  distribution to Euclidean geometry by multiplying it with $S^{2.5}$, where $S$ is the mean flux density of sources in each bin \citep{Windhorst1985}. 

THOR catalogue is 94\% complete above $7\sigma$ \citep{Bihr2016,Wang2018}, whereas, GLOSTAR catalogue is 95\% complete above $7\sigma$ threshold \citep{Medina2019}. In this analysis we choose sources with flux densities above 7$\sigma$ and hence no completeness correction  was applied to the differential source counts.  Resolution bias causes for underestimation of source counts of extended sources in peak flux density selection during source extraction. However, both surveys checked the resolved sources visually and categorized those separately in the final catalogue. We only use the  unresolved/compact sources in this analysis, and hence, this bias is not an issue here. Eddington bias is significant near the detection threshold (5$\sigma$ for these catalogues) due to steep source counts at low flux densities and  the fact that this bias redistributes low flux density sources to high flux bins. However, the differential source counts will not be affected by this bias due to the imposed high flux cut (>7$\sigma$) in this analysis.  Any false detection of observational sidelobe artifacts as real sources will boost the source counts. However, both surveys identified these artifacts by visual inspection of all sources and excluded those from the final catalogue \citep{Wang2018,Medina2019}. 

The Euclidean-normalized differential source counts for THOR and GLOSTAR catalogues are shown in  Fig. \ref{dn_ds}, where the error bars are Poissonian. In the left panel of Fig. \ref{dn_ds}, we compare our findings with two simulated catalogues, the SKA Design Study simulations (SKADS, \citealt{Wilman2008}) and the Tiered Radio Extragalactic Continuum Simulations (T-RECS, \citealt{Bonaldi2019}). The SKADS catalogue spans an area of 100 $\mathrm{deg}^{2}$ with a minimum flux density of 1 $\mu$Jy at 1.4 GHz and also includes four distinct source types:  Fanaroff-Riley Class I (FRI) and Class II (FRII), radio-quiet AGNs (RQQ) and star-forming galaxies (SFGs). The source counts of SFGs, AGNs and RQQ taken from SKADS catalogue are also shown. There are three different settings available in T-RECS simulation for the two main radio source populations: AGNs and SFGs. We choose the `medium' T-RECS catalogue, which covers 25 $\mathrm{deg}^{2}$ with a minimum flux density of 10 nJy at 1.4 GHz and also incorporates the effect of clustering in their simulation \citep{Bonaldi2019}. The population of AGNs and SFGs in T-RECS simulation are also shown in Fig. \ref{dn_ds}.  

Along with simulated radio catalogues we have also compared differential source counts with observed source populations at low-frequency as well as high-frequency in the right panel of Fig. \ref{dn_ds}. The differential source counts from other observations include : the TIFR GMRT Sky
Survey  at 150 MHz (TGSS-ADR1; \citealt{Intema2017}), GaLactic and Extragalactic All-sky MWA survey  at 154 MHz (GLEAM; \citealt{Franzen2019}), BOOTES field at 150 MHz using LOFAR \citep{Williams2016}, Super-CLASS supercluster at 325 MHz with the GMRT \citep{Riseley2016}, ELAIS-N1 field at 400 MHz using uGMRT \citep{Chakraborty2019}, Lockman Hole field at 1.4 GHz with the LOFAR \citep{Prandoni2018}, the 1.4 GHz source counts based on observation with VLA by \citet{Condon1984}, the Phoenix Deep Survey at 1.4 GHz with ATCA \citep{Hopkins2003}, COSMOS field at 3 GHz with VLA \citep{Smolcic2017}. In all cases we have scaled the source counts to 1.4 GHz using a spectral index, $\alpha = -0.8$.

\begin{table} 
\label{Table_1}
\begin{center}
\caption{Best-fit values of amplitude (A) and power-law index ($\gamma$) of w($\theta$) for the unclassified and compact sources in THOR survey and for the all and $z>0.5$ samples in the FIRST survey. The best fitted values for various flux density thresholds are also shown.}
\label{clustering_properties}
\begin{tabular}[\columnwidth]{lccc}
\hline
\hline
Survey & $S_{\mathrm{min}}$ [mJy] & $\mathrm{log_{10}}$(A) & $\gamma$\\
 \hline
       & 0.4 ($7\sigma$) & $-1.38 \pm 0.01$ & $1.60 \pm 0.02$ \\
THOR   & 1         & $-1.71 \pm 0.02$ & $1.79 \pm 0.04$ \\
       & 2         & $-1.78 \pm 0.01$ & $1.88 \pm 0.04$ \\
\hline
FIRST  & all   & $-2.30 \pm 0.76$ & $1.82 \pm 0.02$ \\ 
       & $z>0.5$   & $-1.88 \pm 0.46$ & $2.04 \pm 0.12$ \\
       
\hline
\hline
\end{tabular}
\end{center}
\end{table}

 We have found that the normalized differential source counts of THOR and GLOSTAR survey are in good agreement  with both simulated source models as well as observed source counts. There is a flattening in source population below 1 mJy for THOR catalogue and it closely follows the T-RECS source models. This flattening is attributed to the rise in the source population of SFGs and RQ-AGNs.  However, the contribution of SFGs and RQ-AGNs to the source population below 1 mJy at 1.4 GHz is not well understood and previous observations also find a discrepancy in comparison with SKADS simulation \citep{Smolcic2017, Bonaldi2019}.  It should be noted that for the current analysis only a small part of the GLOSTAR survey catalogue is used as the data analysis and source identification for the full survey region is not complete yet. Also, there is a possibility of non-detection of steep spectrum extragalactic sources  in  higher frequency  GLOSTAR survey (4-5 GHz), which may be the reason for the difference between THOR and GLOSTAR source counts. However, a more detailed comparison will be more appropriate only after the catalogue from the entire survey area is available.
 The Euclidean-normalized differential source counts of these unclassified and compact sources detected in  both surveys are consistent with simulation and observations. This strikingly similar feature confirms that these sources are  of extragalactic origin.  

\begin{figure}
    \centering
 
    \includegraphics[width=\columnwidth,height=3in]{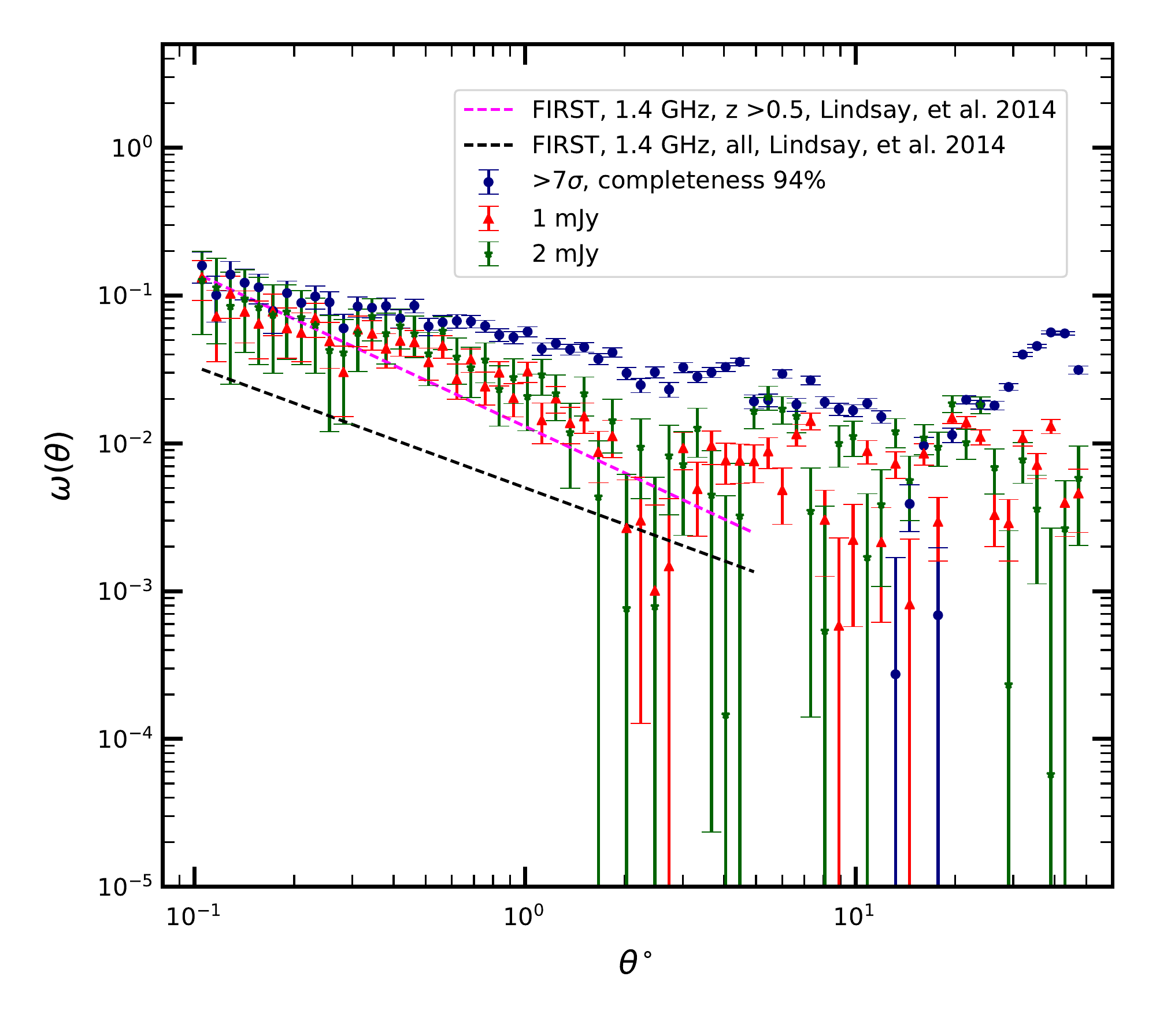}
    \caption{ The two-point angular correlation function of unclassified and unresolved sources detected in THOR survey with flux densities above 0.4 mJy ($7\sigma$), 1 mJy and 2 mJy.Also, the angular clustering of THOR sources  compared with the FIRST survey \citet{Lindsay2014} and plotted in log-scale.}
    \label{w_theta_fig}
\end{figure}


\section{The angular two-point correlation function}

We have also estimated the ATCF of unclassified and compact sources detected in THOR. We have used the LS estimator proposed by \citet{Landy1993}, 
\begin{equation}
    w(\theta) = \frac{DD(\theta) - 2 DR(\theta) + RR(\theta)}{RR(\theta)}
    \label{w_theta}
\end{equation}, 
where DD, DR and RR corresponds to pair counts at a separation of angle $\theta$ for data-data, data-random and random-random catalogue respectively. We have generated a  artificial random catalogue containing a large number of randomly distributed sources across the survey area. However, due to non-uniform noise  across the survey area all sources with different flux densities cannot be detected  across the entire area. In order to incorporate the effect of non-uniform noise in the random catalogue, we have injected 1000 artificial point sources in random positions in each noise map of THOR survey with flux densities  drawn randomly from SKADS 1.4 GHz catalogue. Then we have extracted the  sources following the same criterion as described in \citet{Wang2018} from this simulated map. The extracted source catalogue gives one realization of a artificial random catalogue where the effect of non-uniform noise is also taken care of. We repeat this process until the artificial random catalogue is 20 times the original data catalogue. 

We have used the publicly available code TreeCorr \footnote{\url{https://github.com/rmjarvis/TreeCorr}} \citep{Jarvis2004}  to estimate $w$($\theta$). We have binned the sources between 0.1 deg to 50 deg with bin width of 0.1 deg in log space. We have estimated $w$($\theta$) of the unclassified and compact sources in THOR survey with flux densities above 0.4 mJy (7$\sigma$), 1 mJy and 2 mJy. These flux cuts help us to compare our results with the FIRST survey, which is limited to 1 mJy at 1.4 GHz and also 95\% complete at 2 mJy. We expect that beyond 1 mJy flux density threshold, the point source completeness of THOR is more than 99\%. The ATCF, $w$($\theta$), with Poisson error bars is shown in the Fig. \ref{w_theta_fig}. We have also estimated the `bootstrap' errors \citep{Ling1986} and found that it is larger than the Poisson errors by a factor of two or three in small scales. Note that \citet{Cress1996} have found a similar trait in error estimates for the FIRST survey. 

We have found that for three flux cuts the behavior of $w$($\theta$) is consistent with each other. There is a decrease in correlation with increasing flux density threshold. In Fig. \ref{w_theta_fig}, we have also compared $w$($\theta$) with the FIRST survey \citep{Lindsay2014}. They have analyzed the angular clustering of all the FIRST sources with flux density above 1 mJy at 1.4 GHz and also studied the clustering of sources with redshifts below and above $\textit{z}$ = 0.5. They have found that the high $\textit{z}$ sources are strongly clustered and mainly hosted by massive haloes. \citet{Siewert2019} have found a similar features in angular correlation function for LoTSS-DR1 radio sources.  

We found that at large scales ($\theta > 5^{\circ}$), $w$($\theta$) is dominated by systematics and this work is limited up to this scale.  The noise properties are an issue even at 7$\sigma$ threshold and better understanding of noise distribution across the survey is required. Hence, we limit our analysis upto $5^{\circ}$. We fit the data points in Fig. \ref{w_theta_fig}  between 0.1 deg to 5 deg to a power-law of the form: 
\begin{equation}
    w(\theta) = A (\theta/\mathrm{deg})^{1-\gamma}
\end{equation}

We run Markov Chain Monte Carlo (MCMC) and  Metropolis-Hastings algorithm to estimate the parameters by minimizing the $\chi^{2}$ value.  The best fitted values of A and $\gamma$ for that different flux cuts are mentioned in the Table \ref{Table_1}.  We also estimate the angular correlation with two different subsets of the whole data set and find that the amplitude of clustering is consistent within the errorbars. We find higher flux cuts exhibit smaller correlation amplitude. Also, the clustering amplitude is higher in comparison with the FIRST survey for all sources. However, the angular correlation for 1 mJy flux threshold is in good statistical agreement with  $z > 0.5$ samples of the FIRST survey \citep{Lindsay2014}. This pattern of ATCF suggests that most of these unclassified and compact objects in THOR survey ($\sim 55\%$) may be extragalactic sources at high redshifts. 

The excess correlation  may be due to the additional correlation present in the artificial random catalogue. This can be the result of unaccounted inhomogeneity of completeness (significant near 5$\sigma$ detection threshold) and noise variation across the FoV in generating the random catalogue. However, the artificial catalogue was generated by simulating  sources in the noise plane and then extracting those simulated sources. Hence, we do expect that the variation of noise and completeness across the survey area are taken care of during this procedure. Also, the high flux cuts used in this analysis ensures that variation of completeness do not affect the result  significantly.    

\section{Conclusion}
 THOR and GLOSTAR surveys have a large sample of unclassified sources with unknown origin. Here, we study the possibility that they have properties of  extragalactic population.

First, we have estimated  the differential source counts of unclassified and compact sources in THOR and GLOSTAR survey. There is an excellent agreement of differential source counts with other extragalactic surveys and simulated catalogues. This confirms that these sources are of extragalactic origin. 

 Furthermore, we studied the ATCF of THOR sources using different flux cuts. We found that the clustering amplitude  is higher than  the FIRST survey. However, the angular correlation for sources with flux densities above 1 mJy is in agreement with the clustering of high $z$ sources in the FIRST survey. We also found that as we increase the flux density threshold the clustering amplitude decreases. These features suggest that most of these unclassified THOR and GLOSTAR sources could be  extragalactic with high redshifts.  However, further multi-wavelength (optical, infrared etc) study of these sources are needed to confirm these findings.
 
  This study shows that most of these compact sources detected in different surveys of the Galactic plane are originally of extragalactic origin. So, it is essential to identify or characterize sources detected in these surveys very precisely. There should be one-to-one cross-matching of the sources with other high-frequency catalogues while searching for Galactic compact objects (e.g. compact HII regions, compact components in star forming regions) to avoid possible contamination of extragalactic sources. The ongoing and upcoming surveys with the current or future radio telescopes, like MWA \citep{Tingay2013}, ASKAP \citep{Norris2011}, MeerKAT \citep{Jarvis2016}, SKA \citep{Koopmans2015} etc, should be cautious about this possible contamination in stuying the Galactic objects.\\

{\bf ACKNOWLEDGEMENTS}

We would like to thank the anonymous reviewer  and the scientific editor for suggestions and comments that have helped to improve this work. AC would like to acknowledge DST for providing INSPIRE fellowship. AC would like to thank James J. Condon for providing the 1.4 GHz source counts over private communication. NR acknowledges support from the Max Planck Society through the Max Planck India Partner Group grant. H.B. and Y.W. acknowledge support from the European Research Council under the Horizon 2020 Framework Program via the ERC Consolidator Grant CSF-648505. H.B. further acknowledges support from the Deutsche Forschungsgemeinschaft in the Collab-orative Research Center (SFB 881) “The Milky Way System” (subproject B1). AD would like to acknowledge the support of EMR-II under CSIR No. 03(1461)/19.





\begin{thebibliography}{99}
\bibitem[\protect\citeauthoryear{Beuther et al.}{2016}]{Beuther2016} Beuther H., et al., 2016, A\&A, 595, A32

\bibitem[\protect\citeauthoryear{Bihr et al.}{2016}]{Bihr2016} Bihr S., et al., 2016, A\&A, 588, A97

\bibitem[\protect\citeauthoryear{Bonaldi et al.}{2019}]{Bonaldi2019} Bonaldi A., et al., 2019, MNRAS, 482, 2

\bibitem[\protect\citeauthoryear{Chakraborty et al.}{2019}]{Chakraborty2019} Chakraborty A., et al., 2019, MNRAS, 490, 243

\bibitem[\protect\citeauthoryear{Condon}{1984}]{Condon1984} Condon J.~J., 1984, ApJ, 287, 461


\bibitem[Cress et al.(1996)]{Cress1996} Cress, C.~M., Helfand, D.~J., Becker, R.~H., Gregg, M.~D., White, R.~L.\ 1996.\ 
The Angular Two-Point Correlation Function for the FIRST Radio Survey.\ The Astrophysical Journal 473, 7.

\bibitem[\protect\citeauthoryear{de Zotti et al.}{2010}]{Zotti2010} de Zotti G., Massardi M., Negrello M., Wall J., 2010, A\&ARv, 18, 1

\bibitem[\protect\citeauthoryear{Franzen et al.}{2019}]{Franzen2019} Franzen T.~M.~O., et al., 2019, PASA, 36, e004

\bibitem[\protect\citeauthoryear{Hales et al.}{2012}]{Hales2012} Hales C.~A., Murphy T., Curran J.~R., Middelberg E., Gaensler B.~M., Norris R.~P., 2012, MNRAS, 425, 979

\bibitem[\protect\citeauthoryear{Hopkins et al.}{2003}]{Hopkins2003} Hopkins A.~M., Afonso J., Chan B., Cram L.~E., Georgakakis A., Mobasher B., 2003, AJ, 125, 465

\bibitem[\protect\citeauthoryear{Intema et al.}{2017}]{Intema2017} Intema 
H.~T., Jagannathan P., Mooley K.~P., Frail D.~A., 2017, A\&A,  598, A78 

\bibitem[\protect\citeauthoryear{Jarvis et al.}{2004}]{Jarvis2004} Jarvis M., Bernstein G., Jain B., 2004, MNRAS, 352, 338

\bibitem[\protect\citeauthoryear{Jarvis et al.}{2016}]{Jarvis2016} Jarvis M., et al., 2016, 
    Proceedings of MeerKAT Science: On the Pathway to the SKA. 25-27 May, 2016 Stellenbosch, South Africa (MeerKAT2016). Available at \url{"https://pos.sissa.it/cgi-bin/reader/conf.cgi?confid=277}, id.6.  (arXiv:1709.01901)


\bibitem[\protect\citeauthoryear{Koopmans et al.}{2015}]{Koopmans2015} 
Koopmans L., Pritchard J., Mellema G., et al., 2015, Proceedings of Advancing Astrophysics with the Square Kilometre Array (AASKA14). 9 -13 June, 2014. Giardini Naxos, Italy.Available at: \url{ http://pos.sissa.it/cgi-bin/reader/conf.cgi?confid=215}, id.1.  (arXiv:1505.07568 )

\bibitem[\protect\citeauthoryear{Landy \& Szalay}{1993}]{Landy1993} Landy S.~D., Szalay A.~S., 1993, ApJ, 412, 64

\bibitem[\protect\citeauthoryear{Lindsay et al.}{2014}]{Lindsay2014} Lindsay S.~N., et al., 2014, MNRAS, 440, 1527

\bibitem[\protect\citeauthoryear{Ling, Frenk \& Barrow}{1986}]{Ling1986} Ling E.~N., Frenk C.~S., Barrow J.~D., 1986, MNRAS, 223, 21P

\bibitem[\protect\citeauthoryear{McMullin et al.}{2007}]{McMullin2007} McMullin J.~P., Waters B., Schiebel D., Young W., Golap K., 2007, ASPC..376,  127, ASPC..376

\bibitem[\protect\citeauthoryear{Medina et al.}{2019}]{Medina2019} Medina S.-N.~X., et al., 2019, A\&A, 627, A175

\bibitem[\protect\citeauthoryear{Norris et al.}{2011}]{Norris2011} Norris R.~P., et al., 2011, PASA, 28, 215


\bibitem[\protect\citeauthoryear{Prandoni et al.}{2018}]{Prandoni2018} 
Prandoni I., Guglielmino G., Morganti R., Vaccari M., Maini A., 
R{\"o}ttgering H.~J.~A., Jarvis M.~J., Garrett M.~A., 2018, MNRAS,  481, 
4548

\bibitem[\protect\citeauthoryear{Riseley et al.}{2016}]{Riseley2016} Riseley C.~J., et al., 2016, MNRAS, 462, 917

\bibitem[\protect\citeauthoryear{Siewert et al.}{2019}]{Siewert2019} Siewert T.~M., et al., 2019, arXiv, arXiv:1908.10309

\bibitem[\protect\citeauthoryear{Smol{\v{c}}i{\'c}  et al.}{2017}]{Smolcic2017} Smol{\v{c}}i{\'c} V., et al., 2017, A\&A, 602, A1

\bibitem[\protect\citeauthoryear{Tingay et al.}{2013}]{Tingay2013} Tingay S.~J., et al., 2013, PASA, 30, e007


\bibitem[\protect\citeauthoryear{Wang et al.}{2018}]{Wang2018} Wang Y., et al., 2018, A\&A, 619, A124

\bibitem[\protect\citeauthoryear{Wilman et al.}{2008}]{Wilman2008} Wilman 
R.~J., Miller L., Jarvis M.~J., et al., 2008, MNRAS,  388, 1335

\bibitem[\protect\citeauthoryear{Williams et al.}{2016}]{Williams2016} 
Williams W.~L., van Weeren R.~J., R{\"o}ttgering H.~J.~A., et al., 2016, 
MNRAS,  460, 2385

\bibitem[\protect\citeauthoryear{Windhorst et al.}{1985}]{Windhorst1985} 
Windhorst R.~A., Miley G.~K., Owen F.~N., Kron R.~G., Koo D.~C., 1985, ApJ,  
289, 494 



\end{thebibliography}




\appendix


\bsp	
\label{lastpage}
\end{document}